\RequirePackage{fix-cm}
\documentclass[twocolumn]{svjour3}     
\smartqed 
\usepackage{graphicx}
\usepackage{hyperref}

\begin{document}

\title{Performance evaluation in the reconstruction of 2D images of computed tomography using massively parallel programming CUDA
}


\newcommand{\orcid}[1]{\href{https://orcid.org/#1}{\includegraphics[width=8pt]{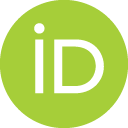}}}

\author{Alexssandro Ferreira Cordeiro \orcid{0000-0002-9493-9398} \and
	 Pedro Luiz de Paula Filho \and Hamilton Pereira da Silva \and Arnaldo Candido Junior \and Edresson Casanova \and Jandrei Sartori Spancerski \orcid{0000-0002-4077-281X}
}


\institute{Department of Computing at the Federal Technological University        of Parana \\
       Brazil Avenue, 4232 \\
       zip code 85884-000 \\
       Medianeira - PR, Brazil
       \email{alexssandrocordeiro@alunos.utfpr.edu.br}      %
}

\date{Received: date / Accepted: date}

\maketitle

\begin{abstract}

Analysis of processing time and similarity of images generated between CPU and GPU architectures and sequential and parallel programming. For image processing a computer with AMD FX-8350 processor and an Nvidia GTX 960 Maxwell GPU was used, along with the CUDAFY library and the programming language C\# with the IDE Visual studio. The results of the comparisons indicate that the form of sequential programming in a CPU generates reliable images at a high custom of time when compared to the forms of parallel programming in CPU and GPU. While parallel programming generates faster results, but with increased noise in the reconstructed image. For data types float a GPU obtained best result with average time equivalent to $\frac{1}{3}$ of the processor, however the data is of type double the parallel CPU approach obtained the best performance. For the float data type, the GPU had the best average time performance, while for the double data type the best average time performance was for the parallel approach CPU. Regarding image quality, the sequential approach obtained similar outputs, while theparallel approaches generated noise in their outputs.

\end{abstract}

\section{Introduction}

The discovery of x-ray by Dr. Wilhelm Conrad Roentgen in 1895 allowed the rise of many studies developed to create non-invasive and highly accurate examinations for patients. In 1972, engineer Godfrey Hounsfield created a computed tomography scanner. He believed that there could be more information on radiographs that could be captured with the film.

Computed Tomography is a high precision non invasive imaging exam, providing a better quality of patient care. CT scan uses x-rays to make the diagnoses and the images are formed through the attenuations that the x-rays suffer when crossing a certain body \cite{trabCorrelato}\cite{trabCorrelato2}.

One of the ways to create 3D computed tomography images is through the interpolation of a stack of 2D images, which requires processing time to reconstruct each image \cite{Interp3D}.

The present study aims to find the method with the highest performance in the delivery of computed tomography exam results, which benefits the patient with a shorter waiting time. It is noteworthy that there was a generation of noise arising from the techniques applied to reduce the time in the reconstruction of 2D images.

In this way, this study is delimited in the process of 2D image reconstruction, focusing on the comparison between the time of image reconstruction and analysis of the quality of these images, through three approaches which will be presented in the methodology section.

\section{Related Works}

\cite{trabCorrelato} demonstrates in his work that the Filtered Backprojection algorithm for reconstruction of 2D tomography images can be massively parallelized and presents a benchmark between an Intel Paragon supercomputer and a Conection Machine CM-5 with dataset based on one image of CT. The algorithm was analyzed for efficiency and speed in Intel Paragon and CM-5. The execution times obtained from the parallelization indicate that at least in the 2D case, overall Intel Paragon delivered better acceleration and efficiency results than the CM-5.

In \cite{trabCorrelato2}, the reconstruction of 2D tomography images demonstrated necessitating a massive parallelization of the algorithm for performance gain. They presented a hybrid approach proposal with parallelization in both CPU and GPU with dataset based on cucumber phantoms. The results demonstrate that the GPU 980 Maxwell used in the study obtained a gain in performance of approximately 5 times, demonstrating the possibility of using GPUs for the reconstruction of electrical impedance tomography images.

In \cite{imagesDifferent}, the reconstruction of 3D images of Tomography using a massively parallel aproach in GPU and sequential approach in a CPU, through the 2D synograms of three different CT images. In the study, is presented two different NVIDIA GPU, Tesla and Fermi architectures were used, which analyzed the quality of the generated images and the performance. It was noticed an acceleration factor between 15 and 85 times when compared to the sequential CPU approach. It was compared the Fermi and Tesla GPUs as well as the quality of the generated images, perceiving noise increment in the images generated by the massively parallel approaches, due to the problem of running in the process of writing-modification-reading of the threads, being necessary the use of computational atomic operations.

\section{Computed tomography}

\begin{figure}[htbp]
	\centering{\includegraphics[width=8cm,clip]{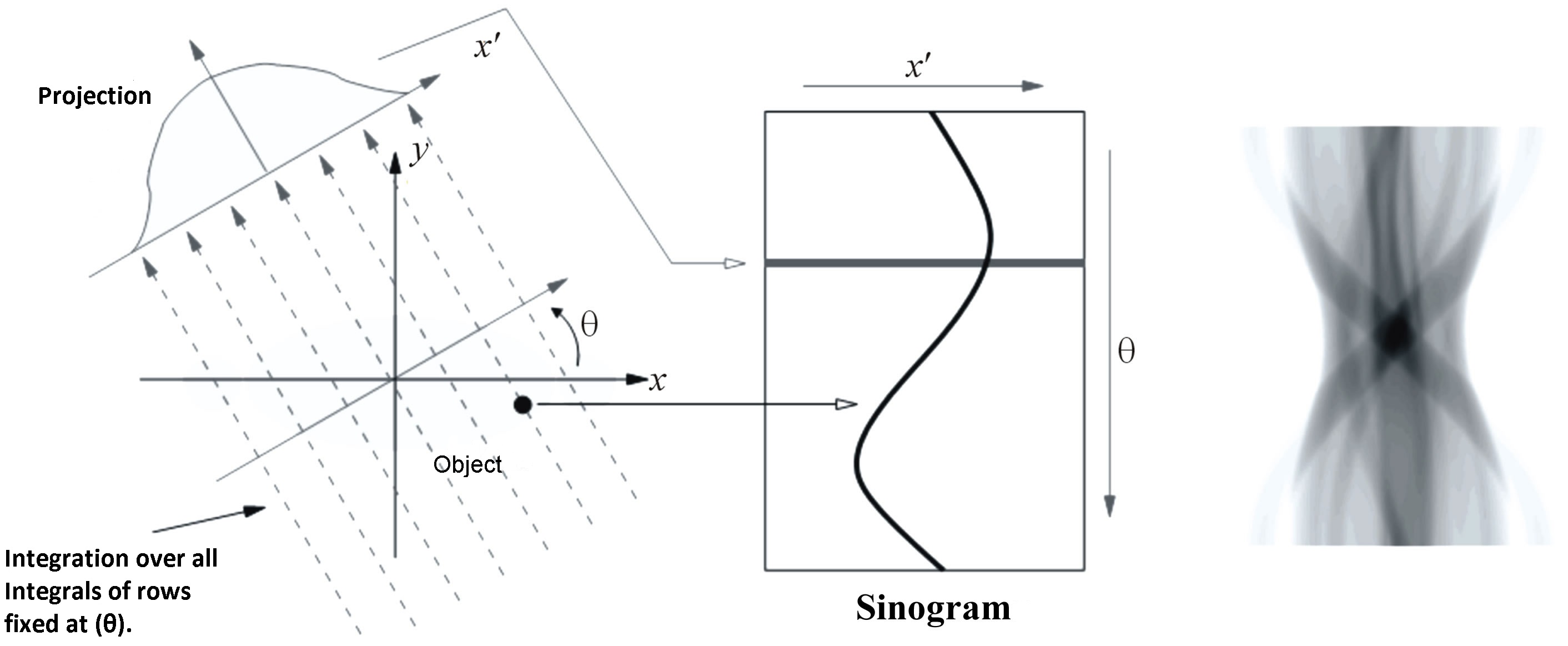}}
	\caption{Illustration of projection and sinogram .} 
	\label{fig:sinogramaadp} 
\end{figure}

Computed tomography is performed using the Johann Radon algorithm or Radon transform, also known as sinograms, which aims to reconstruct images of sections or slices of a body from measurements of attenuations that the x-ray suffered when crossing the body at a certain angle $\theta$, as illustrated in [\cite{Asl2013}, Figure~\ref{fig:sinogramaadp} (adapted)].


The law that studies the physical process of the interaction of an x-ray beam with matter is the Lambert Beer law according to \cite{lambertbeer} and is seen in \ref{eqlambertbeer}.

\begin{equation}
I = I_0 \cdot e^{-\mu \cdot d}
\label{eqlambertbeer}
\end{equation}

\begin{figure}[htbp]
	\centering{\includegraphics[width=5cm,clip]{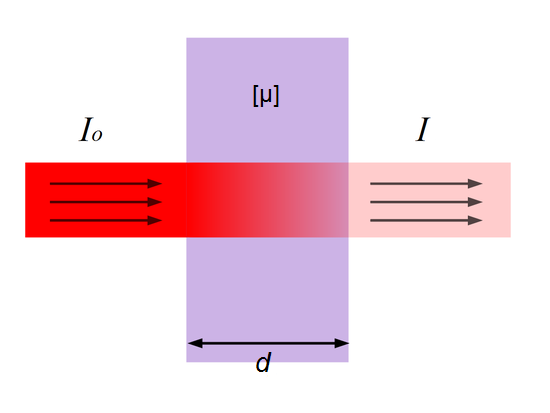}}
	\caption{Lambert Beer law illustration.} 
	\label{fig:lambertbeer} 
\end{figure}

Where I is the intensity of the beam after passing through the body; {$I_0$} is the initial beam intensity and (-{$\mu \cdot$ \textit{d}}) is the product of the attenuation coefficient~($\mu$) with the thickness of the body~(\textit{d}) hit by the beam, as illustrated in [\cite{kinectLambertBeert}, Figure~\ref{fig:lambertbeer} (adapted)].

Thus, when performing a CT scan, it is possible to obtain a matrix with the attenuation values of the x-ray beams in a two-dimensional space {$ f (x, y) $}, representing slices of an object.

\begin{figure}[!htb]
	\centering{\includegraphics[width=8cm,clip]{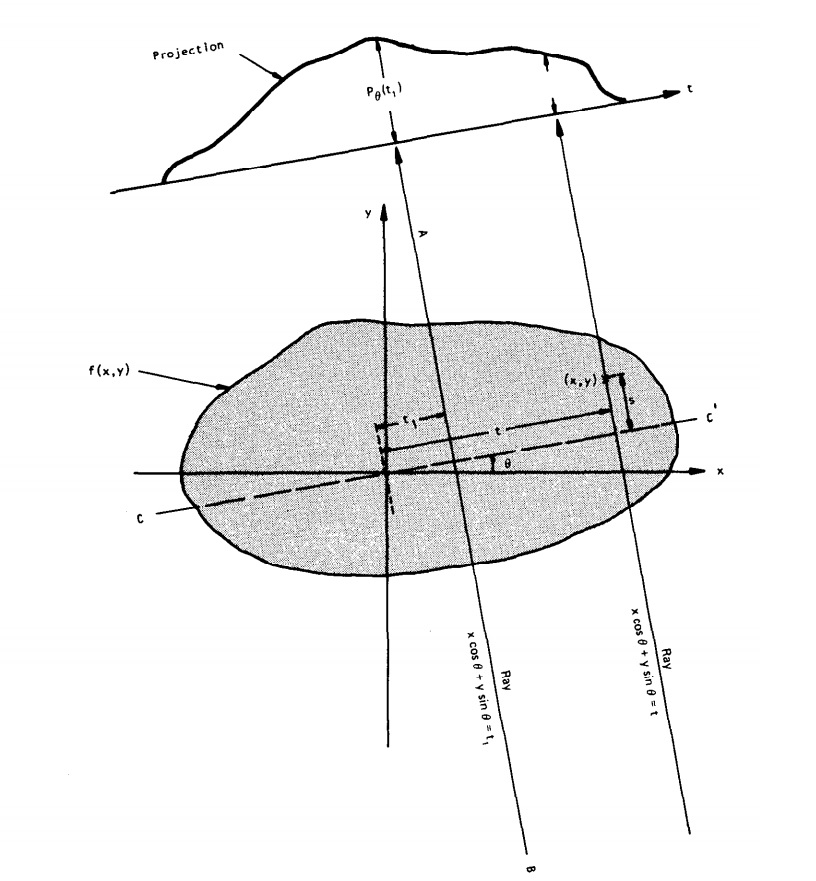}}
	\caption{One-Dimensional projection.} 
	\label{fig:projuni} 
\end{figure}

Knowing that the x-ray travels in a straight line as can be observed in the [\cite{principiosimagenstomografia}, Figure~\ref{fig:projuni}], where each line is in function of f(x,y) and integrated by the parameters ({$\theta , t$}), in this way each beam of x-ray can be described by equation~($x \cdot cos(\theta)+y \cdot sin(\theta)=t$).

Applying the filtered backpropagation (FBP) algorithm and the inverse Radon transform, the equation \ref{eqgeralcasoreal} is obtained.

\begin{equation}
f(x,y) \ = \ \sum_{i}^{k} \ Q\theta_i (t) [\ x \cdot \ cos(\theta_i) \ + \ y \cdot \ sin(\theta_i)]
\label{eqgeralcasoreal}
\end{equation}

Where $Q_\theta(t) = P_\theta(t) \ast H_\xi(t)$, representing the filter convolution over the original projection. O algoritmo de reconstrução das imagens 2D de tomografia computadorizada se utiliza desta equação.

\section{Benchmark}

Computed tomography images were created from data obtained through the website of the Technical University of Denmark\footnote{www.bme.elektro.dtu.dk/31545/?ct\_data/ct\_data.html},
these data are part of the project \textit{The Visible Human}\footnote{www.nlm.nih.gov/research/visible/visible\_human.html} of the National Library of Medicine of the United States and are composed of files with numerical data referring to the attenuations that the x-ray suffered during the examination.

In this study we used a computer with AMD\footnote{www.amd.com/pt/products/cpu/fx-8350} FX-8350 processor with 8 cores and 4.0 GHz, 8GB of memory DDR3 kingston\footnote{www.kingston.com/us/gaming/hyperx-fury-ddr3} HyperX Fury with 1600Mhz, a SSD kingston UV400 240GB and an NVIDIA GTX 960 video card
integrated with Maxwell architecture GPU's, 1024 CUDA cores and 4GB of GDDR5 memory.

The study took into account three main approaches wich are sequential CPU, CPU with thread utilization and GPU accelerated computing using CUDA technology. In this way, the study analyze which approach has better performance without loss of image quality, were used in reconstructions of images data types of simple precision float 32 bits and double 64 bits precision.

We have reconstructed 13 different images, each image was reconstructed 10 times, adding 130 reconstructed images per approach, each approach was performed 5 times, thus adding a total of 1950 reconstructions per data type. thus obtaining a total of 3900 image reconstructions for this study.

The images were analyzed using the Peak Signal to Noise Ratio (PSNR) and Structural Similarity Index (SSIM) algorithms to ensure the quality of the generated images.

According to \cite{psnrSsimImagem}, PSNR is an estimate of the image reconstructed with the original through the pixel differences, its algorithm is given by \ref{eqPSNR}.

\begin{equation}
PSNR= 20*\log_{10}(\frac{MAX_{i}}{\sqrt{MSE}})
\label{eqPSNR}
\end{equation}

In the PSNR algorithm, the greater the result given in decibels (dB), the more similar the images are, if the result is indeterminate or $\infty$, it means that the images are the same because they obtained zero in the denominator.

In agreement with \cite{SSIMImage}, SSIM is an algorithm used to verify the similarity of images from loss of luminance, correlation, distortion and contrast distortion, its is seen in \ref{eqSSIM}.

\begin{equation}
SSIM(x,y) = \frac{(2\mu_x \mu_y + C_1)(2 \sigma_{xy} + C_2)}{(\mu_x^2 + \mu_y^2 + C_1)(\sigma_x^2 + \sigma_y^2 + C_2)}
\label{eqSSIM}
\end{equation}

For SSIM, if the result is 1, the images are the same, if it is close to 1, the images are similar.

\section{Results and discussion}

\begin{figure}[!htb]
	\begin{center}
		\includegraphics[height=1.5cm]{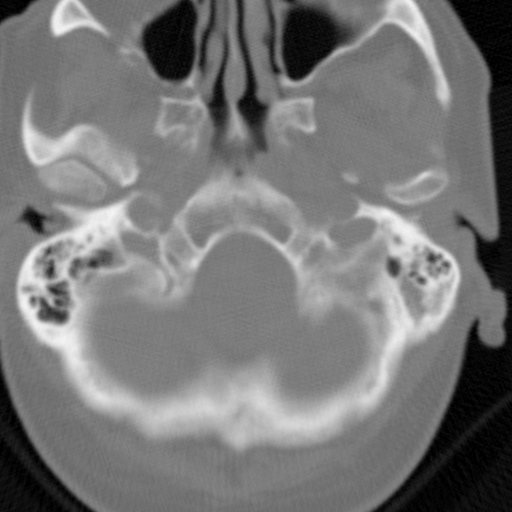} \quad
		\includegraphics[height=1.5cm]{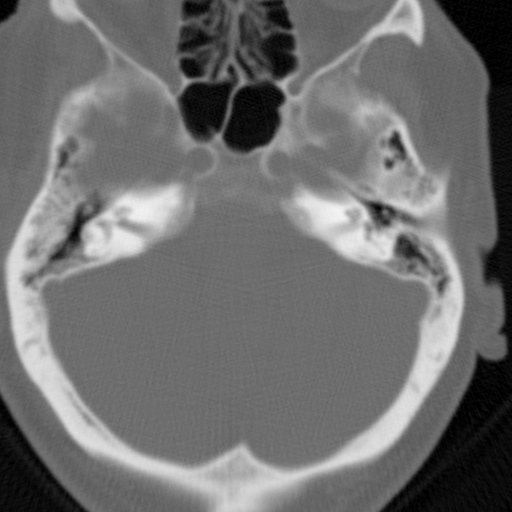}
		\includegraphics[height=1.5cm]{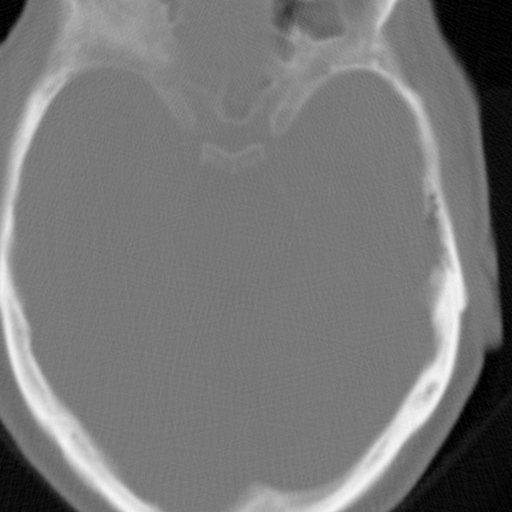}
		\includegraphics[height=1.5cm]{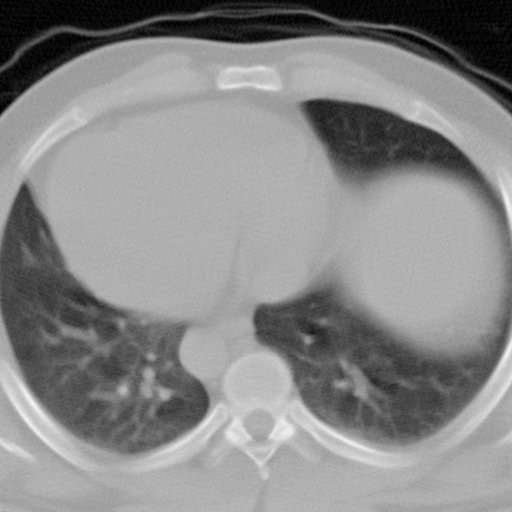}
		\includegraphics[height=1.5cm]{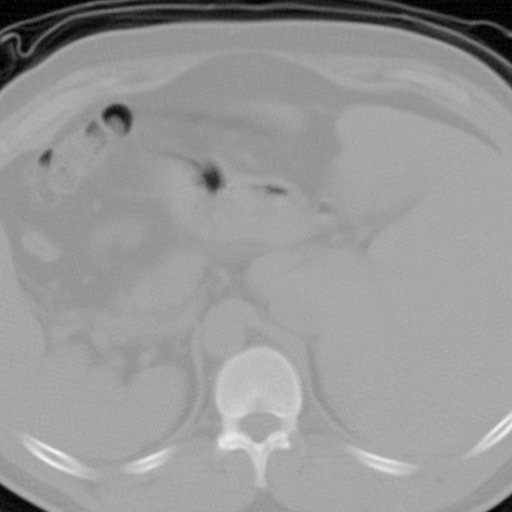}
		\includegraphics[height=1.5cm]{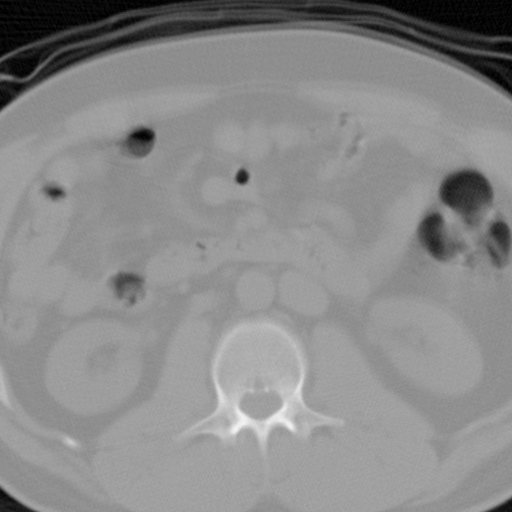}
		\includegraphics[height=1.5cm]{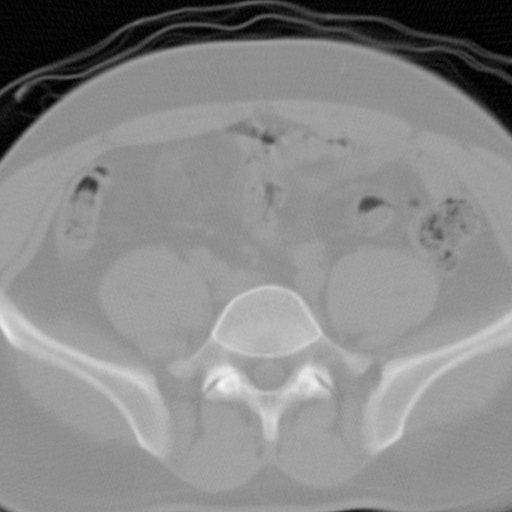}
		\includegraphics[height=1.5cm]{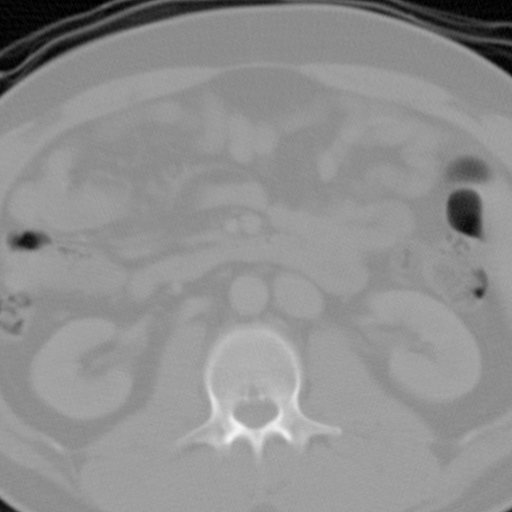}
		\includegraphics[height=1.5cm]{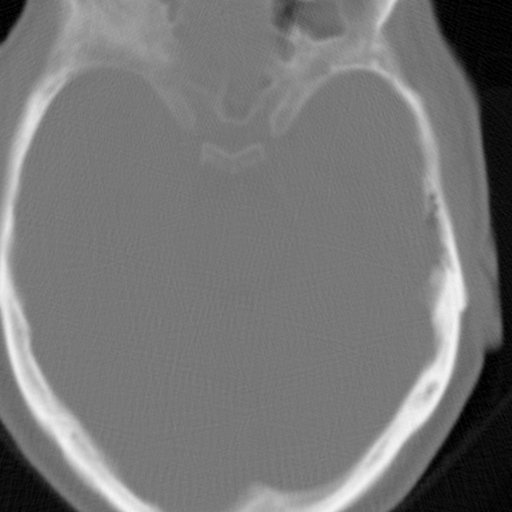}
		\includegraphics[height=1.5cm]{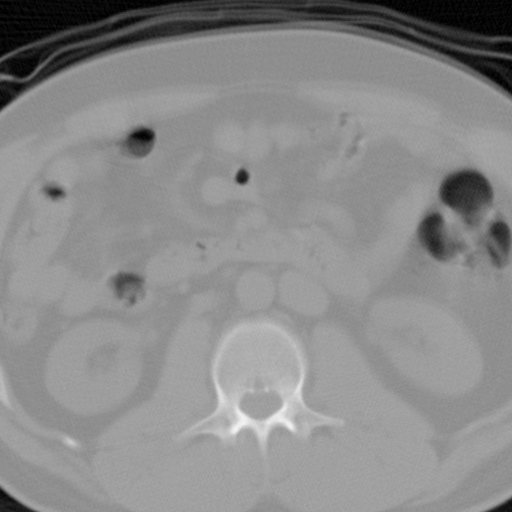}
		\includegraphics[height=1.5cm]{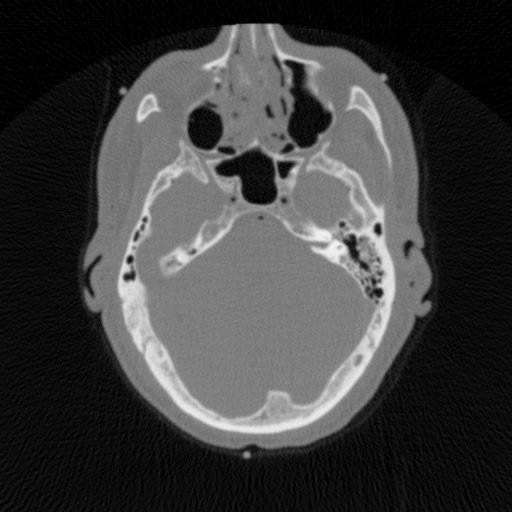}
		\includegraphics[height=1.5cm]{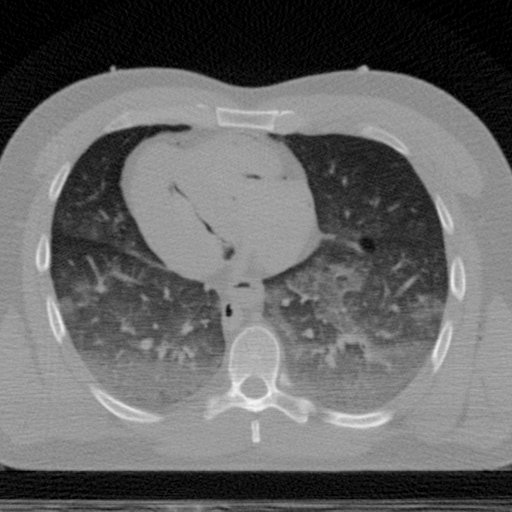}
		\includegraphics[height=1.5cm]{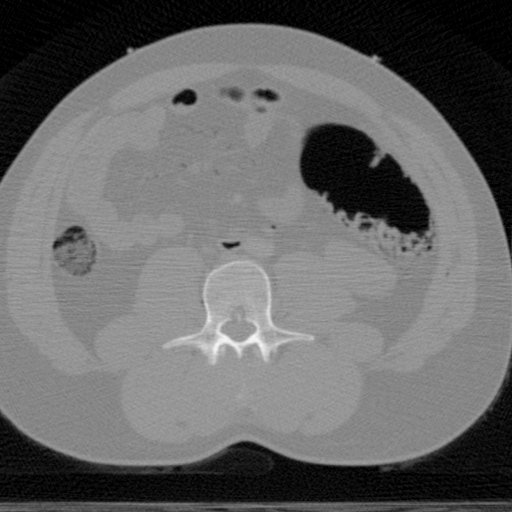}
		\caption{2D images rebuilt by the application.}
		\label{fig:imagensrec} 
	\end{center}
\end{figure}

The images that make up the dataset used in the bechmark are illustrated in Figure~\ref{fig:imagensrec}.

\begin{figure}[htbp]
	\centerline{\includegraphics[height=5cm,clip]{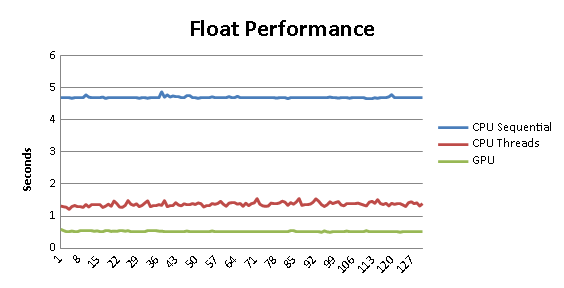}}
	\caption{Graphics with the performance of 2D image reconstructions using the 32-bit float data type.}
	\label{fig:graficosFloat} 
\end{figure}

After executing the bechmark, the results obtained with the float data types can be observed as illustrated in the graph of Figure~ \ref{fig:graficosFloat}.

The average GPU time in the reconstruction of a 2D image was 0.52 seconds, the CPU reconstruction with threads was 1.37 seconds and the sequential CPU was 4.69 seconds. In this way, the performance of the GPU in the reconstruction of the images of computed tomography 2D with the use of the type float data is highlighted.

In discussion of the results found regarding time in float data type, for the presented hardware configuration, the GPU obtained an average of approximately $\frac{1}{3}$ of the parallel CPU time, due to the number of cores of CUDA on the GPU, compared to the number of CPU cores. A similar fact occurs when comparing theads CPU approaches with the sequential CPU, which has a determining factor in the number of cores used in processing, with the performance of the CPU threads being less than $\frac{1}{3}$ of the sequential CPU time.

Referring to the quality of the images, it can be presented from the tables \ref{tab:qualidadeimgPSNRfloat} and \ref{tab:qualidadeimgSSIMfloat} that the images are similar in each approach.

The Table \ref{tab:qualidadeimgPSNRfloat} demonstrates the similarity of the images generated between the approaches, using the float data type, through the PSNR image quality analysis algorithm.

\begin{table}[htbp]
	\caption{Comparison between 2D images, reconstructed with the float data type, using the PSNR algorithm, values	in ``dB''.} 
	\begin{center}
		\begin{tabular}{|c|c|c|}
			\hline \hline
			CPU - CPU Thread & CPU - GPU & CPU Thread - GPU\\
			\hline
			49,9449 & 84,8574 & 54,0245 \\
			\hline
			55,3524 & 45,4621 & 51,5397 \\
			\hline
			$\infty$ & 50,8809 & 53,711 \\
			\hline
			$\infty$ & 48,0042 & 43,5866 \\
			\hline
			36,4884 & 85,8053 & 36,7921 \\
			\hline
			48,0352 & 85,8053 & 36,7921 \\
			\hline
			40,1446 & 87,7834 & 46,2455 \\
			\hline
			62,37 & 85,401 & 85,401 \\
			\hline
			$\infty$ & 53,2306 & 53,711 \\
			\hline
			48,0352 & 48,0342 & 84,1872 \\
			\hline
			36,582 & 38,8628 & 51,1872 \\
			\hline
			40,5224 & 49,7154 & 52,1694 \\
			\hline
			50,5089 & 50,5096 & 50,6431 \\
			\hline \hline
		\end{tabular}
		\label{tab:qualidadeimgPSNRfloat}
	\end{center}
\end{table}

The Table \ref{tab:qualidadeimgSSIMfloat} demonstrates the similarity of the images generated between the approaches, using the float data type, through the SSIM image quality analysis algorithm.

\begin{table}[htbp]
	\caption{Comparison between 2D images, reconstructed with the floating data type, using the SSIM algorithm.} 
	\begin{center}
		\begin{tabular}{|c|c|c|}
			\hline \hline
			CPU - CPU Thread & CPU - GPU & CPU Thread - GPU\\
			\hline
			0,998344 & 0,999999 & 0,998599 \\
			\hline
			0,998802 & 0,998586 & 0,99804 \\
			\hline
			1 & 0,997344 & 0,997926 \\
			\hline
			1 & 0,999255 & 0,998736 \\
			\hline
			0,997545 & 0,999998 & 0,997214 \\
			\hline
			0,999549 & 0,999998 & 0,999549 \\
			\hline
			0,997301 & 0,999999 & 0,998282 \\
			\hline
			0,999593 & 0,999998 & 0,999998 \\
			\hline
			1 & 0,997772 & 0,997926 \\
			\hline
			0,999549 & 0,99548 & 0,999999 \\
			\hline
			0,996765 & 0,999548 & 0,999999 \\
			\hline
			0,998777 & 0,999306 & 0,998677 \\
			\hline
			0,994856 & 0,998455 & 0,998398 \\
			\hline \hline
		\end{tabular}
		\label{tab:qualidadeimgSSIMfloat}
	\end{center}
\end{table}

\begin{figure}[htbp]
	\centerline{\includegraphics[height=5cm,clip]{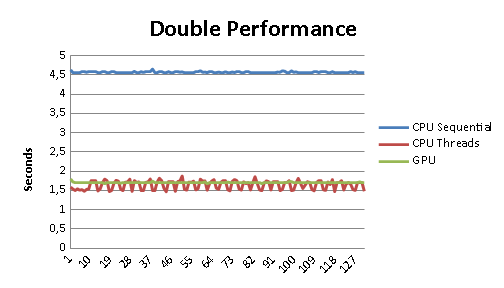}}
	\caption{Graphics with the performance of 2D image reconstructions using the 64-bit double data type.}
	\label{fig:graficosDouble} 
\end{figure}

Regarding the results obtained with the double data types, they can be observed by means of the graphic of the Figure~\ref{fig:graficosDouble}.

The average CPU time with threads in this approach was 1.66 seconds, a GPU got 1.93 seconds, and the sequential CPU averaged 4.57 seconds for better performance compared to a float version. Regarding GPU performance with double data type, according to \cite{gpudoublevsfloat}, a bandwidth, an increase in the number of bytes to read from 4 to 8, and architecture of CUDA processors, influence GPU performance in this way , obtaining higher values than the CPU with Thread.

In this experiment, the CPU with Threads obtained better performance in the average time compared to the GPU, this result being given by the fact that the GPU Maxwell architecture prioritizes the float data type, obtaining a better output and getting more gigaflops compared to the double data type \cite{gpudoubleperformance}.

Regarding the quality of the generated images, it is possible to perceive through the tables \ref{tab:qualidadeimgPSNRdouble} and \ref{tab:qualidadeimgSSIMdouble} that the images are similar.

The table \ref{tab:qualidadeimgPSNRdouble}, demonstrates the similarity of the generated images between the approaches, using the double data type, through the PSNR image quality analysis algorithm.

\begin{table}[htbp]
	\caption{Comparison between 2D images, reconstructed with the double data type, using the PSNR algorithm, values in ``dB''.} 
	\begin{center}
		\begin{tabular}{|c|c|c|}
			\hline \hline
			CPU - CPU Thread & CPU - GPU & CPU Thread - GPU\\
			\hline
			49,9439 & 85,5985 & 49,9435 \\
			\hline
			$\infty$ & 42,8479 & 86,3701 \\
			\hline
			53,712 & 53,7102 & 53,7102 \\
			\hline
			48,0048 & 48,0037 & 48,0038 \\
			\hline
			39,7314 & 86,1353 & 30,6757 \\
			\hline
			48,0346 & 87,3092 & 49,2355 \\
			\hline
			40,1449 & 86,4925 & 40,1446 \\
			\hline
			48,2527 & 86,4925 & 86,4925 \\
			\hline
			$\infty$ & 87,0195 & 53,7102 \\
			\hline
			42,9966 & 87,3092 & 42,9965 \\
			\hline
			36,5819 & 37,4668 & 53,5696 \\
			\hline
			40,5226 & 84,3744 & 43,5129 \\
			\hline
			50,509 & 85,9125 & 44,7184 \\
			\hline \hline
		\end{tabular}
		\label{tab:qualidadeimgPSNRdouble}
	\end{center}
\end{table}

The table \ref{tab:qualidadeimgSSIMdouble}, demonstrates the similarity of the images generated between the approaches, using the double data type, through the SSIM image quality analysis algorithm.

\begin{table}[htbp]
	\caption{Comparison between 2D images, reconstructed with the double data type, using the SSIM algorithm.} 
	\begin{center}
		\begin{tabular}{|c|c|c|}
			\hline \hline
			CPU - CPU Thread & CPU - GPU & CPU Thread - GPU\\
			\hline
			0,998345 & 0,999999 & 0,998345 \\
			\hline
			1 & 0,997993 & 0,999999 \\
			\hline
			0,997927 & 0,997926 & 0,997929 \\
			\hline
			0,999256 & 0,999255 & 0,999256 \\
			\hline
			0,997 & 0,999998 & 0,996507 \\
			\hline
			0,99955 & 0,999998 & 0,998029 \\
			\hline
			0,997301 & 0,999998 & 0,997301 \\
			\hline
			0,999152 & 0,999999 & 0,999999 \\
			\hline
			1 & 0,999999 & 0,997926 \\
			\hline
			0,997928 & 0,999998 & 0,997928 \\
			\hline
			0,996765 & 0,997236 & 0,9990001 \\
			\hline
			0,998777 & 0,999999 & 0,999202 \\
			\hline
			0,998457 & 0,999998 & 0,998029 \\
			\hline \hline
		\end{tabular}
		\label{tab:qualidadeimgSSIMdouble}
	\end{center}
\end{table}

It is noteworthy that with each new reconstruction, a different value of PSNR and SSIM can be obtained, as discussed in the work of \cite{imagesDifferent}, the authors define that this is due to the process of writing-modification-reading of the threads, being necessary the use of computational atomic operations to eliminate noise, but causing loss of performance. This does not happen when performed in the sequential CPU approach, where the results obtained by PSNR and SSIM demonstrate that the images are the same.

\section{Conclusion}

According to the results obtained, it is possible to conclude that the images generated in each approach are similar above 99\%, but when performing the interpolation of a 2D batch of tomography images, unreliable results can be generated in the 3D volumetric images, given to the accumulation of noise generated both in the reconstruction and in the interpolation of the image. 

Regarding the double data type, in order to guarantee better performance, it is necessary to make a prior analysis in the architecture of the hardware used before guaranteeing which methodology performs better.


It is also concluded that the GPU has better performance in the float data type when compared to a conventional CPU and that the CPU using threads obtained better performance in the double data type. Thus, the performance item depends on both the data type and the CPU/GPU architecture used in the experiment.

As a continuation of this research, we intend to develop the reconstruction of 3D images of computed tomography using the algorithm of Feldkamp, Davis and Kress (FDK) according to~\cite{FutureWork}, which uses the filtered backprojection algorithm in the reconstruction of three-dimensional images obtained by projections of a cone beam of x-ray. To accelerate the reconstruction of 3D images, we intend to use a massively parallel programming with CUDA and after the reconstructions, to perform an analysis of the noise added in the images in comparison to a sequential programming in the CPU, it is worth mentioning that a 3D image is obtained from of multiple 2D images.



\end{document}